\documentclass[aps,pre,twocolumn,groupedaddress,showpacs]{revtex4}
\usepackage{graphicx}
\usepackage{amssymb}
\usepackage{array}

\begin{document}


\title{Temperature-pressure scaling for air-fluidized grains on approaches to Point~J}


\author{L. J. Daniels$^1$, T. K. Haxton$^{2}$, N. Xu$^{3}$, A. J. Liu$^1$, and D. J. Durian$^1$}
\affiliation{$^{1}$Department of Physics and Astronomy, University
of Pennsylvania, Philadelphia, PA 19104-6396, USA}
\affiliation{$^{2}$Molecular Foundry, Lawrence Berkeley National Laboratory, Berkeley, CA 94720, USA}
\affiliation{$^{3}$Department of Physics, University of Science and Technology of China, Hefei, Anhui, 230026, P.R. China}


\date{\today}

\begin{abstract}
We present experiments on a monolayer of air-fluidized beads in which a jamming transition is approached by increasing pressure, increasing packing fraction, and decreasing kinetic energy. This is accomplished, along with a noninvasive measurement of pressure, by tilting the system and examining behavior vs depth. We construct an equation of state and analyze relaxation time vs effective temperature. By making time and effective temperature dimensionless using factors of pressure, bead size, and bead mass, we obtain a good collapse of the data but to a functional form that differs from that of thermal hard-sphere systems. The relaxation time appears to diverge only as the effective temperature to pressure ratio goes to zero.
\end{abstract}

%
%


\maketitle



The relaxation time for amorphous liquids can grow unbearably long when the temperature is lowered~\cite{NagelChem, DebStill01}.  It can also grow when the pressure is increased, although this is more difficult to study experimentally~\cite{DanielGiles02, RolandRPP05, WinMenon06}.  Similarly, the relaxation time for colloidal suspensions can exceed experimentalists' patience when the packing fraction or pressure is increased~\cite{LCip, Mattson, VoigtmannPoon06}.  In both the thermal and colloidal glass transitions, the particles appear to develop a fixed set of neighbors and the bulk medium appears to become mechanically rigid.  It was recently suggested that these two glass transitions are manifestations of the same phenomenon for the system of thermal hard spheres~\cite{Ning}.  In such a system, dimensional analysis suggests that the relaxation time $\tau$, made dimensionless as $\tau (P \sigma^{d-2}/m)^{1/2}$ by pressure $P$, the sphere diameter $\sigma$, and the sphere mass $m$, must depend only on the dimensionless ratio $T/P \sigma^d$, where $d$ is the dimensionality and the Boltzmann constant is set to unity.  Thus, the dimensionless relaxation time increases in exactly the same way whether $T$ is lowered or $P$ is raised. 

Although no system behaves exactly like hard spheres, Medina-Noyola and coworkers showed that there should be ``dynamical equivalence," so that soft spheres behave as hard spheres with a smaller diameter~\cite{MMN03, MMN11}.  Indeed, it was  found that particles with a variety of finite-ranged repulsive interaction potentials exhibited collapse of dimensionless relaxation time with $T/p \sigma^d$ as long as the pressure was low, so that $P \ll \varepsilon / \sigma^d$, where $\varepsilon$ is the interaction energy scale~\cite{Ning}. These results suggest that real systems might exhibit the temperature-pressure scaling expected for hard spheres; however, this has not been tested by experiment.  For hard-sphere colloids, this is not possible for a single system because the packing fraction $\phi$, or equivalently the pressure $P$, is the only control parameter; temperature is bounded by the freezing and boiling points of the solvent and therefore cannot be varied appreciably.  For molecular liquids, scaling is not expected to hold where van der Waals attractions are appreciable compared to the pressure; in this regime the scaling must be modified to account for the mean-field effect of the attractions \cite{VoigtmannPoon06}.

Here we describe experiments on a granular monolayer of bidisperse beads, subjected to random in-plane forcing by turbulence in a uniform up-flow of air \cite{Abate2006, Abate2007, Glotzer2007, AbatePRL08}.  In this system, jamming may be approached by variation of either air flow rate or bead density~\cite{Abate2006}.   The thermal energy of these beads is negligible compared with their kinetic energy due to airflow.  However, it has been shown that there is a  well-defined effective temperature, $T_{\rm eff}$, that corresponds equivalently to the kinetic energy, the ratio of diffusivity to mobility, and the energy of an embedded harmonic oscillator~\cite{AbatePRL08}.  Here, we push the concept of effective temperature further by asking whether temperature-pressure scaling is obeyed using $T_{\rm eff}$.  

Indeed, we find reasonable collapse with $T_{\rm eff}/P \sigma^2$ of both the equation of state and relaxation time, even though our system is far from equilibrium.  However, the functional dependence on the temperature-pressure ratio is different from that in a true thermal system.  These results suggest that while the effective thermal glass transition and the colloidal glass transition remain the same phenomenon even for these driven hard spheres, the glass transition dynamics differ from those of equilibrium hard spheres.


Fluidization and imaging procedures follow Refs.~\cite{Abate2006, Abate2007, Glotzer2007, AbatePRL08, Ojha2005}.  The granular medium is a 1:1 mixture of 796 steel bearings, with diameters $\sigma=0.397$~cm and $1.4\sigma$, which uniformly fill the $(14.9~{\rm cm})^2$ sample square to 67\% projected area fraction when the system is level.  The superficial airflow speed is 700~cm/s, which is sub-levitating and corresponds to Re=2000.  To measure pressure, we tilt the entire apparatus to induce a component of gravity along the monolayer and determine the mass per unit length that is above a given depth. When tilted by angle $\theta$, ranging here from $0.18^\circ$ to $0.90^\circ$, the grains are fluidized near the top but become progressively jammed with increasing depth $z$ below the top edge.  Video data are collected at 120~frames per second, typically for 20~minutes, and are analyzed in strips of width $\Delta z = 0.84$~cm.  At each depth we determine the time average of the packing fraction $\phi$, the pressure $P=m_zg\sin\theta/L$ where $m_z$ is the mass of all beads between $z$ and the top edge, and the mean-squared displacement.  From the latter we deduce both the granular effective temperature $T_{\rm eff}$ as the average bead kinetic energy, and the relaxation time $\tau$.  Here, $\tau$ is defined as the time needed for the root-mean-squared displacement to equal $\sigma$.  Technical details are available on-line \cite{SOM}.

\begin{figure}
\includegraphics[width=3in]{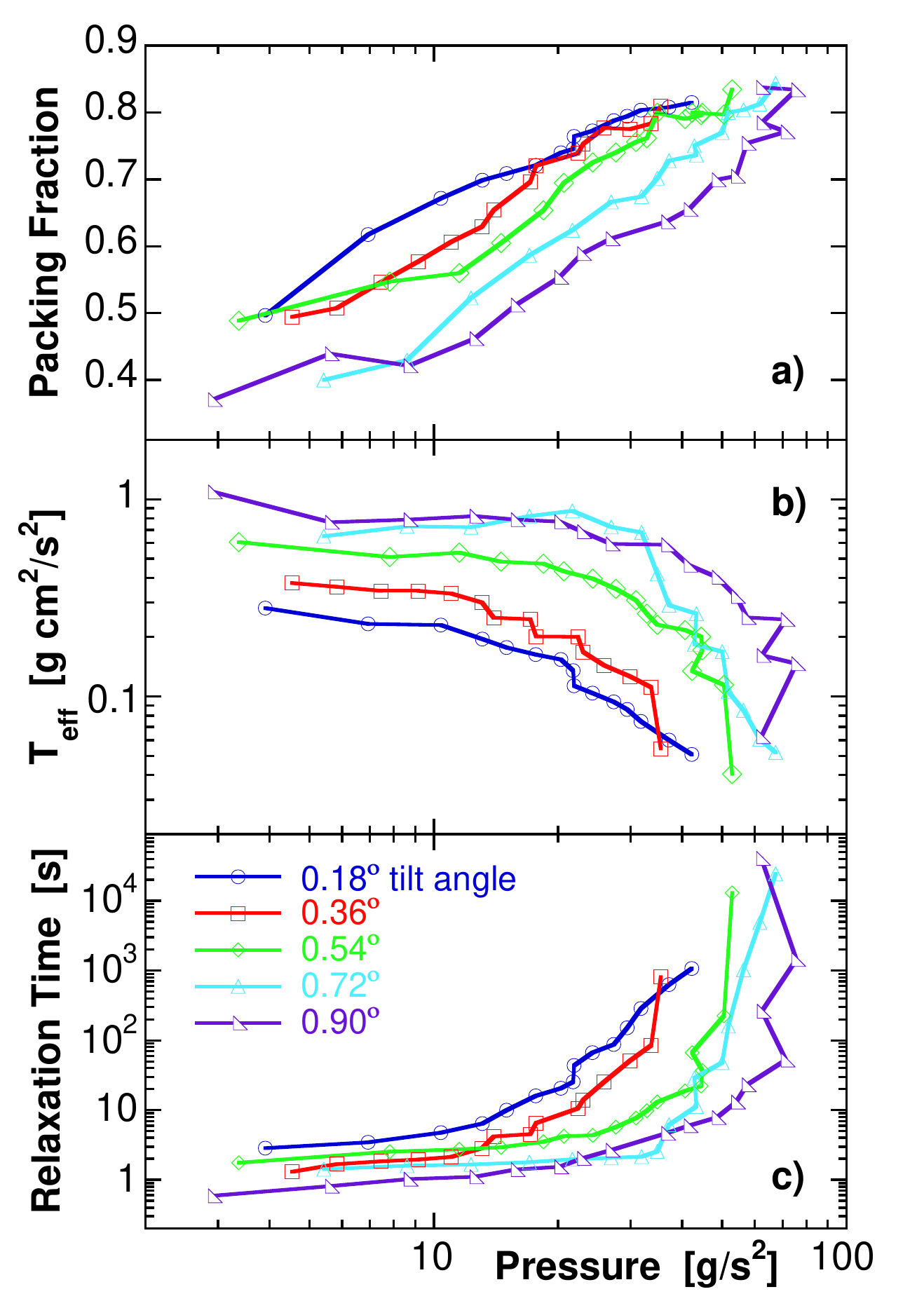}
\caption{(Color online)(tphivp.pdf) Projected-area packing fraction, effective temperature, and relaxation time, all plotted parametrically versus pressure for different tilt angles as labeled.}\label{TPHIVP}
\end{figure}

The final results for $\phi$, $T_{\rm eff}$, and $\tau$ are plotted vs pressure $P$ in Fig.~\ref{TPHIVP} for five different tilt angles $\theta$.  Each point corresponds to a different depth, such that the pressure is zero at the top edge and increases with depth.  Fig.~\ref{TPHIVP} shows that as $P$ increases towards the bottom, jamming is approached in that both $\phi$ and $\tau$ increase while $T_{\rm eff}$ decreases.  Note that at a given $P$ there is a substantial range in $\phi$, $T_{\rm eff}$, and $\tau$ values for the different tilt angles; therefore, any collapse achieved by temperature-pressure scaling will have significance.

\begin{figure}
\includegraphics[width=3in]{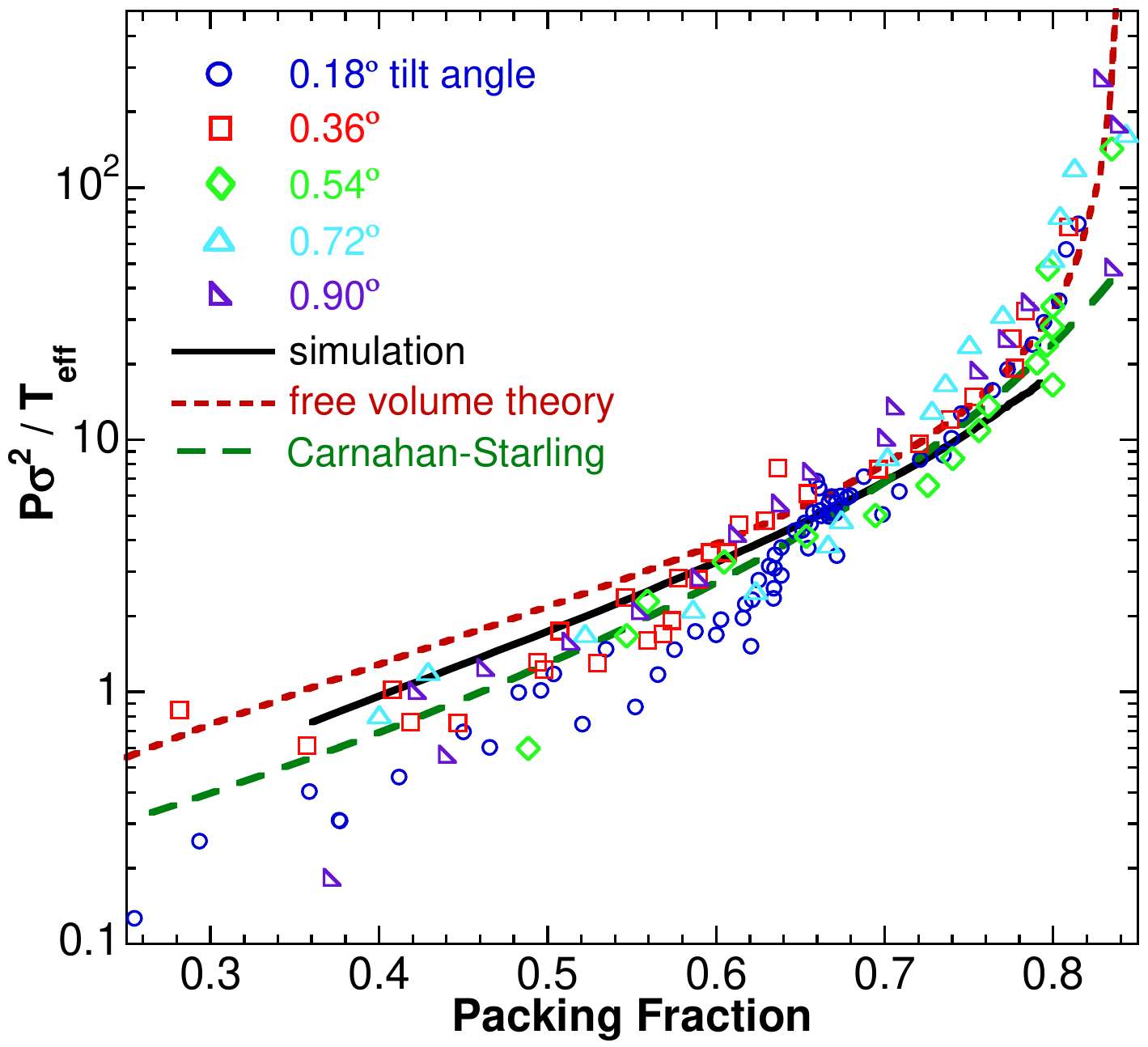}
\caption{(Color online)(eqofstateall.pdf) Equation of state for pressure divided by temperature, vs packing fraction, where $\sigma$ is the small bead diameter.  The solid curve is for a two-dimensional simulation of disks interacting via hard-core repulsion. The short and long dashed curves are fits to free-volume and Carnahan-Starling forms, respectively.}\label{EQofST}
\end{figure}

The first quantity we consider is the equation of state, which is independent of the bead dynamics.  Thus we form the dimensionless combination $P\sigma^2/T_{\rm eff}$  and plot it vs $\phi$ in Fig.~\ref{EQofST}.   This causes the raw data shown in Fig.~\ref{TPHIVP}a-b to collapse reasonably well onto a single curve, with scatter that is random in tilt angle.  As expected the dimensionless pressure is low for small $\phi$, grows with increasing $\phi$, and appears to diverge as $\phi\rightarrow\phi_c$ where $\phi_c\approx0.84$ corresponds to random close packing for hard spheres in two dimensions~\cite{SalBook}.  For comparison, we show fits to two forms: free-volume theory, $P\sigma^2/T_{\rm eff}\propto \phi/[1-(\phi/\phi_c)^{1/2}]$, and Carnahan-Starling~\cite{Carnahan, Randy}.  Both give a reasonable description, but the former fits better near jamming and the latter fits better away from jamming.  

We note that the collapse of $P\sigma^2/T_{\rm eff}$ with $\phi$  is remarkable, given the nonequilibrium nature of the system.  Our result shows that the effective temperature gives rise to a well-defined equation of state, underscoring the conclusion of Ref.~\cite{AbatePRL08} that the air-fluidized beads have an effective temperature with thermodynamic meaning.  

The measured equation of state may also be compared with simulation results for a thermal system.  For this, event-driven molecular dynamics are performed in the microcanonical ensemble for a 1:1 bidisperse mixture of 1024 hard-core disks with the same diameter and mass ratios as in the experiment.   The results for $P\sigma^2/T$ are plotted as a solid curve vs $\phi$ in Fig.~\ref{EQofST}.   In comparison with experimental data, the simulated equation of state is slightly high for $\phi<0.5$ and does not diverge rapidly enough for $\phi>0.8$.  One possible source of this discrepancy is that the air-mediated bead-bead repulsion is strong or long-ranged, as measured previously for two beads alone \cite{Ojha2005}.  However, the interactions are based on turbulent wakes and hence are not pair-wise additive; plus, a similar bead-boundary interaction is washed out for many-bead systems.  Furthermore, if air-mediated interactions (pairwise-additive or not) were significant, we would not expect to find collapse, which should only occur in the hard-sphere limit.  Thus we speculate that the nonequilbrium nature of the experimental system is responsible for the discrepancy with the simulation.  The driven system has a well-defined equation of state that just happens to differ slightly from the equilibrium equation of state.



\begin{figure}
\includegraphics[width=3in]{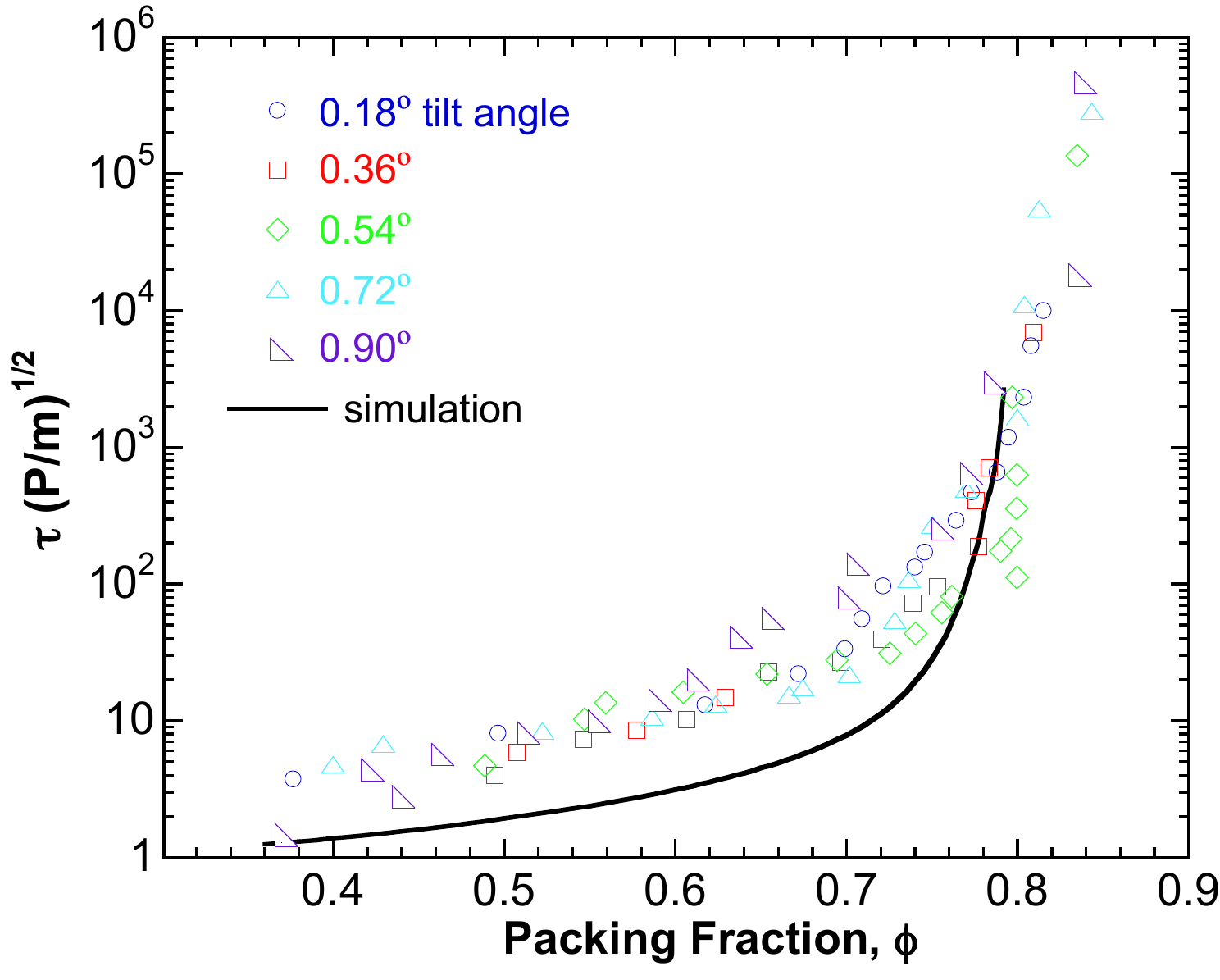}
\caption{(Color online)(trelaxvphiA.pdf) Scaled relaxation time versus packing fraction, where $P$ is pressure and $m$ is the small bead mass, for different tilt angles as labeled.  The solid curve is for a two-dimensional simulation of disks interacting via hard-core repulsion.}\label{tauvphi}
\end{figure}

Next we consider the relaxation time, $\tau$, which is the key dynamical quantity specifying the extent to which the system is jammed.  Since the Reynolds number is large and the dynamics are collisional, the viscosity of air is not relevant for setting the time scale.  Thus we form the dimensionless combination $\tau(P/m)^{1/2}$, as in Ref.~\cite{Ning}, and plot it vs packing fraction in Fig.~\ref{tauvphi}.  Data for the different tilt angles all collapse reasonably well onto a single curve.  Again, the deviation from collapse is not systematic in tilt angle.  Note that $\tau(P/m)^{1/2}$ decreases towards a number of order unity at low $\phi$ far from jamming, reinforcing the conclusion that $\tau(P/m)^{1/2}$ is indeed the correct dimensionless relaxation time to consider.  Above $\phi=0.7$, the relaxation time grows at an ever increasing rate and appears to diverges as $\phi\rightarrow\phi_c$ as expected.  The simulation results for equilibrium hard spheres, in comparison, are lower at small volume fractions and increase more steeply at high volume fractions.

\begin{figure}
\includegraphics[width=3in]{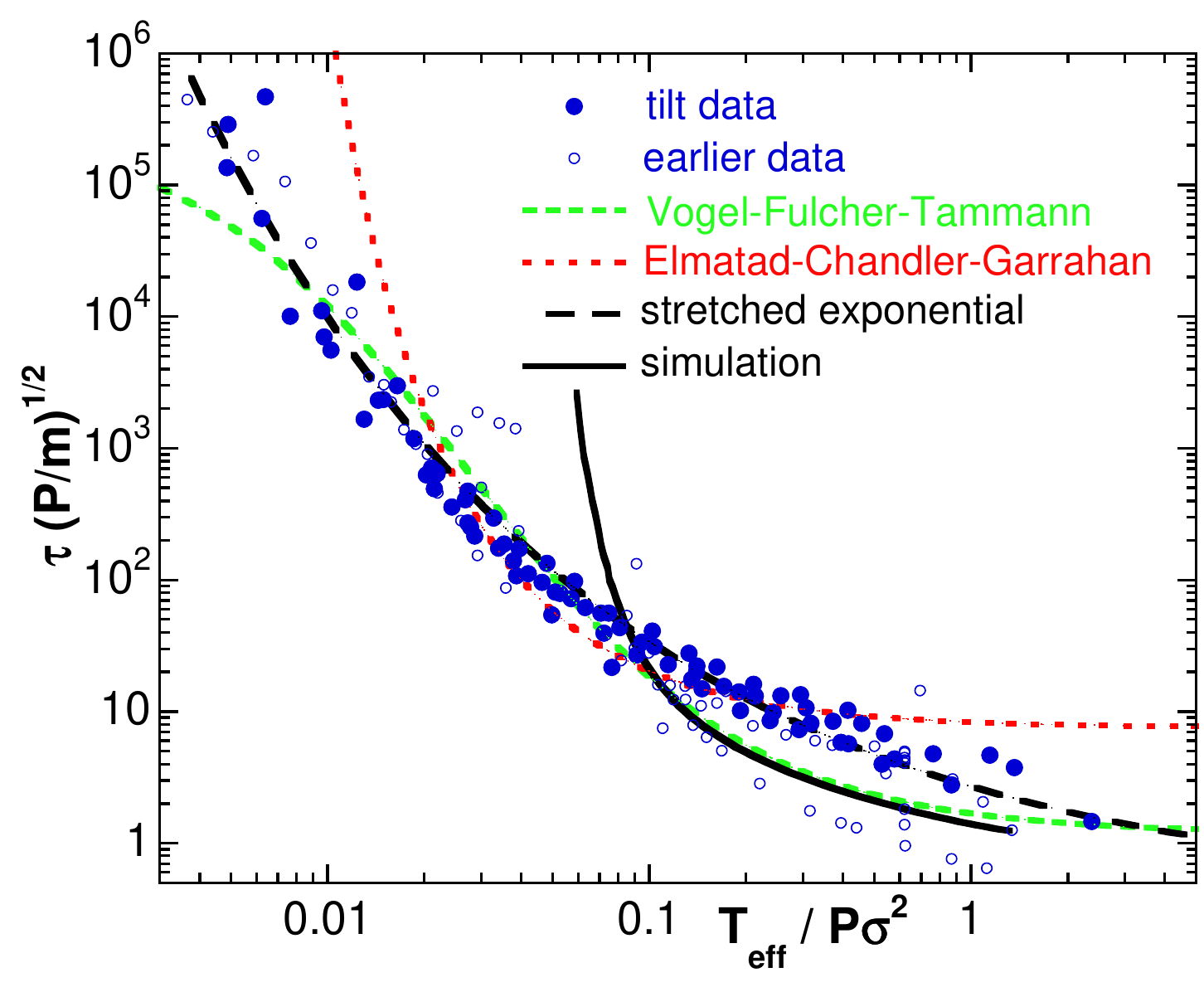}
\caption{(Color online)(tauvT.pdf) Scaling plot of relaxation time vs temperature, where $P$ is pressure and $m$ and $\sigma$ are respectively the mass and diameter of the small beads.  The solid circles are the same data shown in Fig.~\ref{tauvphi}.  The open circles are earlier data, compiled from Refs.~\cite{Abate2006,Abate2007,Glotzer2007,AbatePRL08} and with $T_{\rm eff}/P\sigma^2$ taken according to Fig.~\ref{EQofST} and the known packing fractions.  The solid curve is for a two-dimensional simulation of disks interacting via hard-core repulsion, while the other three curves are fits given by Eq.~(\ref{relax}) as labeled.}\label{tauvT}
\end{figure}

We also plot the scaled relaxation time $\tau(P/m)^{1/2}$ versus scaled temperature $T_{\rm eff}/P\sigma^2$ in Fig.~\ref{tauvT}.  There we also include data from earlier experiments \cite{Abate2006, Abate2007, Glotzer2007, AbatePRL08}, where $T_{\rm eff}/P\sigma^2$ values were obtained by interpolating the measured equation of state in Fig.~\ref{EQofST} to the desired packing fractions.  The good agreement between prior and current data shows that analyzing the video data in narrow strips does not introduce unwanted artifacts.  It also shows that the bead sizes and masses, which are all different except for Refs.~\cite{Abate2007, Glotzer2007}, do not noticeably affect the collapse.  The common behavior of all data, for over two decades in scaled temperature and over five decades in relaxation time, is quite different from the simulation results for equilibrium hard spheres.  In particular, the experimental relaxation times appear to diverge only in the limit $T_{\rm eff}/P\sigma^2\rightarrow 0$, while the simulated relaxation times appear to diverge more rapidly.

In the on-line supplement \cite{SOM} we also show that good collapse is found for the size of dynamical heterogeneities versus the dimensionless relaxation time\cite{LucaDHbook, BerthierP11},  both for air-fluidized beads \cite{Abate2007, Glotzer2007, AbatePRL08} as well as for two other systems on approach to jamming \cite{katsuragi10, KerstinDH}.

The Fig.~\ref{tauvT} scaling plot of relaxation time versus temperature allows comparison with well-known functions for thermal systems.  For this we display fits to the Vogel-Fulcher-Tammann (VFT) and Elmatad-Chandler-Garrahan (ECG) \cite{ECG} forms, as well as a stretched exponential, respectively
\begin{equation}
 \tau(P/m)^{1/2} \propto \cases{ 	\exp[A/(x-x_o)], \cr
 							\exp[B(1/x-1/x_1)^2], \cr
							\exp[C/x^a], \cr}
\label{relax}
\end{equation}
where $x=T_{\rm eff}/P\sigma^2$ and the other quantities are fitting parameters.  The VFT form traditionally accounts for divergence of relaxation time at a nonzero temperature.  But here the best fit gives a proportionality constant of $1.3\pm0.2$, $A=0.36\pm0.02$, and a negative critical temperature $x_o=-0.029\pm0.005$.  The latter causes the fit to roll over to a constant at $T_{\rm eff}/P\sigma^2$ goes to zero, which is not a physical feature supported by the data.  The best fit to the ECG form gives a proportionality constant of $0.0035\pm0.0025$, $B=0.00032\pm0.00002$, and $x_1=-0.00625\pm0.00015$.  This diverges at zero temperature, and gives a better fit than VFT; however, the divergence is too fast.  Overall, the best fit is a stretched exponential with proportionality constant $0.40\pm0.05$, $C=2.0\pm0.6$, and stretching exponent $a=0.35\pm0.04$.  Thus it appears that the data diverge only at zero temperature.

In the case where $T_{\rm eff}$ is well-defined so that different definitions yield the same result, one would expect collapse with $T_{\rm eff}/P \sigma^2$ with the same functional form as in equilibrium.  Indeed, simulations of thermal hard spheres under shear~\cite{Haxton11} support this expectation.  It is surprising that the relaxation time for our driven system shows temperature-pressure scaling but with a functional form that is different from that in equilibrium.  The scaling collapse of relaxation time in our system suggests that the effective thermal glass transition, observed by lowering $T_{\rm eff}$, is equivalent to the ``colloidal" glass transition, observed by varying $P$.  However, the difference in functional form suggests that the glass transitions of the driven system are somewhat different from those of the equilibrium system.  In particular, for equilibrium hard spheres, the value of $T/P \sigma^d$ at which the relaxation time diverges is ambiguous; the VFT form, which diverges at a nonzero value of $T/P \sigma^d$, and the ECG form, which diverges at $T/P \sigma^d=0$, both fit simulation data equally well~\cite{Ning}.  It is still not known whether equilibrium hard spheres have a thermodynamic glass transition (corresponding to a divergence at nonzero $T/P \sigma^d$).  But for the driven granular hard spheres, the answer seems much more clear: The relaxation time appears to diverge at $T/P \sigma^d=0$.  This has a special significance because it corresponds to the zero-temperature jamming transition of spheres, Point~J~\cite{OHernPRE03, LiuNagelARCMP10}.  Our results therefore imply that glassy dynamics in air-fluidized grains are controlled not by a thermodynamic glass transition, but by Point~J.

This work was supported by the NSF through grants DMR-0704147 (LJD, DJD) and DMR-0520020 (TKH), and by the DOE Office of Basic Energy Sciences, Division of Materials Sciences and Engineering under Award DE-FG02-05ER46199 (AJL, NX).

\bibliography{References_pressure}

\begin{thebibliography}{28}
\expandafter\ifx\csname natexlab\endcsname\relax\def\natexlab#1{#1}\fi
\expandafter\ifx\csname bibnamefont\endcsname\relax
  \def\bibnamefont#1{#1}\fi
\expandafter\ifx\csname bibfnamefont\endcsname\relax
  \def\bibfnamefont#1{#1}\fi
\expandafter\ifx\csname citenamefont\endcsname\relax
  \def\citenamefont#1{#1}\fi
\expandafter\ifx\csname url\endcsname\relax
  \def\url#1{\texttt{#1}}\fi
\expandafter\ifx\csname urlprefix\endcsname\relax\def\urlprefix{URL }\fi
\providecommand{\bibinfo}[2]{#2}
\providecommand{\eprint}[2][]{\url{#2}}

\bibitem[{\citenamefont{Ediger et~al.}(1996)\citenamefont{Ediger, Angell, and
  Nagel}}]{NagelChem}
\bibinfo{author}{\bibfnamefont{M.~D.} \bibnamefont{Ediger}},
  \bibinfo{author}{\bibfnamefont{C.~A.} \bibnamefont{Angell}},
  \bibnamefont{and} \bibinfo{author}{\bibfnamefont{S.~R.} \bibnamefont{Nagel}},
  \bibinfo{journal}{J. Phys. Chem.} \textbf{\bibinfo{volume}{100}},
  \bibinfo{pages}{13200} (\bibinfo{year}{1996}).

\bibitem[{\citenamefont{Debenedetti and Stillinger}(2001)}]{DebStill01}
\bibinfo{author}{\bibfnamefont{P.~G.} \bibnamefont{Debenedetti}}
  \bibnamefont{and} \bibinfo{author}{\bibfnamefont{F.~H.}
  \bibnamefont{Stillinger}}, \bibinfo{journal}{Nature}
  \textbf{\bibinfo{volume}{410}}, \bibinfo{pages}{259} (\bibinfo{year}{2001}).

\bibitem[{\citenamefont{Alba-Simionesco
  et~al.}(2002)\citenamefont{Alba-Simionesco, Kivelson, and
  Tarjus}}]{DanielGiles02}
\bibinfo{author}{\bibfnamefont{C.}~\bibnamefont{Alba-Simionesco}},
  \bibinfo{author}{\bibfnamefont{D.}~\bibnamefont{Kivelson}}, \bibnamefont{and}
  \bibinfo{author}{\bibfnamefont{G.}~\bibnamefont{Tarjus}},
  \bibinfo{journal}{J. Chem. Phys.} \textbf{\bibinfo{volume}{116}},
  \bibinfo{pages}{5033} (\bibinfo{year}{2002}).

\bibitem[{\citenamefont{Roland et~al.}(2005)\citenamefont{Roland,
  Hensel-Bielowka, Paluch, and Casalini}}]{RolandRPP05}
\bibinfo{author}{\bibfnamefont{C.~M.} \bibnamefont{Roland}},
  \bibinfo{author}{\bibfnamefont{S.}~\bibnamefont{Hensel-Bielowka}},
  \bibinfo{author}{\bibfnamefont{M.}~\bibnamefont{Paluch}}, \bibnamefont{and}
  \bibinfo{author}{\bibfnamefont{R.}~\bibnamefont{Casalini}},
  \bibinfo{journal}{Rept. Prog. Phys.} \textbf{\bibinfo{volume}{68}},
  \bibinfo{pages}{1405} (\bibinfo{year}{2005}).

\bibitem[{\citenamefont{Win and Menon}(2006)}]{WinMenon06}
\bibinfo{author}{\bibfnamefont{K.~Z.} \bibnamefont{Win}} \bibnamefont{and}
  \bibinfo{author}{\bibfnamefont{N.}~\bibnamefont{Menon}},
  \bibinfo{journal}{Phys. Rev. E} \textbf{\bibinfo{volume}{73}},
  \bibinfo{pages}{040501} (\bibinfo{year}{2006}).

\bibitem[{\citenamefont{Cipelletti and Ramos}(2005)}]{LCip}
\bibinfo{author}{\bibfnamefont{L.}~\bibnamefont{Cipelletti}} \bibnamefont{and}
  \bibinfo{author}{\bibfnamefont{L.}~\bibnamefont{Ramos}}, \bibinfo{journal}{J.
  Phys. Cond. Mat.} \textbf{\bibinfo{volume}{17}}, \bibinfo{pages}{R253}
  (\bibinfo{year}{2005}).

\bibitem[{\citenamefont{Mattsson et~al.}(2009)\citenamefont{Mattsson, Wyss,
  Fernandez-Nieves, Miyazaki, Hu, Reichman, and Weitz}}]{Mattson}
\bibinfo{author}{\bibfnamefont{J.}~\bibnamefont{Mattsson}},
  \bibinfo{author}{\bibfnamefont{H.~M.} \bibnamefont{Wyss}},
  \bibinfo{author}{\bibfnamefont{A.}~\bibnamefont{Fernandez-Nieves}},
  \bibinfo{author}{\bibfnamefont{K.}~\bibnamefont{Miyazaki}},
  \bibinfo{author}{\bibfnamefont{Z.}~\bibnamefont{Hu}},
  \bibinfo{author}{\bibfnamefont{D.~R.} \bibnamefont{Reichman}},
  \bibnamefont{and} \bibinfo{author}{\bibfnamefont{D.~A.} \bibnamefont{Weitz}},
  \bibinfo{journal}{Nature} \textbf{\bibinfo{volume}{462}}, \bibinfo{pages}{83}
  (\bibinfo{year}{2009}).

\bibitem[{\citenamefont{Voigtmann and Poon}(2006)}]{VoigtmannPoon06}
\bibinfo{author}{\bibfnamefont{T.}~\bibnamefont{Voigtmann}} \bibnamefont{and}
  \bibinfo{author}{\bibfnamefont{W.~C.~K.} \bibnamefont{Poon}},
  \bibinfo{journal}{J. Phys.-Cond. Matt.} \textbf{\bibinfo{volume}{18}},
  \bibinfo{pages}{L465} (\bibinfo{year}{2006}).

\bibitem[{\citenamefont{Xu et~al.}(2009)\citenamefont{Xu, Haxton, Liu, and
  Nagel}}]{Ning}
\bibinfo{author}{\bibfnamefont{N.}~\bibnamefont{Xu}},
  \bibinfo{author}{\bibfnamefont{T.~K.} \bibnamefont{Haxton}},
  \bibinfo{author}{\bibfnamefont{A.~J.} \bibnamefont{Liu}}, \bibnamefont{and}
  \bibinfo{author}{\bibfnamefont{S.~R.} \bibnamefont{Nagel}},
  \bibinfo{journal}{Phys. Rev. Lett.} \textbf{\bibinfo{volume}{103}},
  \bibinfo{pages}{245701} (\bibinfo{year}{2009}).

\bibitem[{\citenamefont{de~J.~Guevara-Rodr\'iguez and
  Medina-Noyola}(2003)}]{MMN03}
\bibinfo{author}{\bibfnamefont{F.}~\bibnamefont{de~J.~Guevara-Rodr\'iguez}}
  \bibnamefont{and}
  \bibinfo{author}{\bibfnamefont{M.}~\bibnamefont{Medina-Noyola}},
  \bibinfo{journal}{Phys. Rev. E} \textbf{\bibinfo{volume}{68}},
  \bibinfo{pages}{011405} (\bibinfo{year}{2003}).

\bibitem[{\citenamefont{Ram\'irez-Gonz\'alez
  et~al.}(2011)\citenamefont{Ram\'irez-Gonz\'alez, L\'opez-Flores, Acu\~na
  Campa, and Medina-Noyola}}]{MMN11}
\bibinfo{author}{\bibfnamefont{P.~E.} \bibnamefont{Ram\'irez-Gonz\'alez}},
  \bibinfo{author}{\bibfnamefont{L.}~\bibnamefont{L\'opez-Flores}},
  \bibinfo{author}{\bibfnamefont{H.}~\bibnamefont{Acu\~na Campa}},
  \bibnamefont{and}
  \bibinfo{author}{\bibfnamefont{M.}~\bibnamefont{Medina-Noyola}},
  \bibinfo{journal}{Phys. Rev. Lett.} \textbf{\bibinfo{volume}{107}},
  \bibinfo{pages}{155701} (\bibinfo{year}{2011}).

\bibitem[{\citenamefont{Abate and Durian}(2006)}]{Abate2006}
\bibinfo{author}{\bibfnamefont{A.~R.} \bibnamefont{Abate}} \bibnamefont{and}
  \bibinfo{author}{\bibfnamefont{D.~J.} \bibnamefont{Durian}},
  \bibinfo{journal}{Phys. Rev. E} \textbf{\bibinfo{volume}{74}},
  \bibinfo{pages}{031308} (\bibinfo{year}{2006}).

\bibitem[{\citenamefont{Abate and Durian}(2007)}]{Abate2007}
\bibinfo{author}{\bibfnamefont{A.~R.} \bibnamefont{Abate}} \bibnamefont{and}
  \bibinfo{author}{\bibfnamefont{D.~J.} \bibnamefont{Durian}},
  \bibinfo{journal}{Phys. Rev. E} \textbf{\bibinfo{volume}{76}},
  \bibinfo{pages}{021306} (\bibinfo{year}{2007}).

\bibitem[{\citenamefont{Keys et~al.}(2007)\citenamefont{Keys, Abate, Glotzer,
  and Durian}}]{Glotzer2007}
\bibinfo{author}{\bibfnamefont{A.~S.} \bibnamefont{Keys}},
  \bibinfo{author}{\bibfnamefont{A.~R.} \bibnamefont{Abate}},
  \bibinfo{author}{\bibfnamefont{S.~C.} \bibnamefont{Glotzer}},
  \bibnamefont{and} \bibinfo{author}{\bibfnamefont{D.~J.}
  \bibnamefont{Durian}}, \bibinfo{journal}{Nature Phys.}
  \textbf{\bibinfo{volume}{3}}, \bibinfo{pages}{260} (\bibinfo{year}{2007}).

\bibitem[{\citenamefont{Abate and Durian}(2008)}]{AbatePRL08}
\bibinfo{author}{\bibfnamefont{A.~R.} \bibnamefont{Abate}} \bibnamefont{and}
  \bibinfo{author}{\bibfnamefont{D.~J.} \bibnamefont{Durian}},
  \bibinfo{journal}{Phys. Rev. Lett.} \textbf{\bibinfo{volume}{101}},
  \bibinfo{pages}{245701} (\bibinfo{year}{2008}).

\bibitem[{\citenamefont{Ojha et~al.}(2005)\citenamefont{Ojha, Abate, and
  Durian}}]{Ojha2005}
\bibinfo{author}{\bibfnamefont{R.~P.} \bibnamefont{Ojha}},
  \bibinfo{author}{\bibfnamefont{A.~R.} \bibnamefont{Abate}}, \bibnamefont{and}
  \bibinfo{author}{\bibfnamefont{D.~J.} \bibnamefont{Durian}},
  \bibinfo{journal}{Phys. Rev. E} \textbf{\bibinfo{volume}{71}},
  \bibinfo{pages}{016313} (\bibinfo{year}{2005}).

\bibitem[{SOM()}]{SOM}
\bibinfo{note}{See supplementary material at http://\ldots.}

\bibitem[{\citenamefont{Torquato}(2001)}]{SalBook}
\bibinfo{author}{\bibfnamefont{S.}~\bibnamefont{Torquato}},
  \emph{\bibinfo{title}{Random Heterogeneous Materials: microstructure and
  macroscopic properties}} (\bibinfo{publisher}{Springer},
  \bibinfo{address}{New York}, \bibinfo{year}{2001}).

\bibitem[{\citenamefont{Carnahan and Starling}(1969)}]{Carnahan}
\bibinfo{author}{\bibfnamefont{N.~F.} \bibnamefont{Carnahan}} \bibnamefont{and}
  \bibinfo{author}{\bibfnamefont{K.~E.} \bibnamefont{Starling}},
  \bibinfo{journal}{J. Chem. Phys.} \textbf{\bibinfo{volume}{51}},
  \bibinfo{pages}{635} (\bibinfo{year}{1969}).

\bibitem[{\citenamefont{Kamien and Liu}(2007)}]{Randy}
\bibinfo{author}{\bibfnamefont{R.~D.} \bibnamefont{Kamien}} \bibnamefont{and}
  \bibinfo{author}{\bibfnamefont{A.~J.} \bibnamefont{Liu}},
  \bibinfo{journal}{Phys. Rev. Lett.} \textbf{\bibinfo{volume}{99}},
  \bibinfo{pages}{155501} (\bibinfo{year}{2007}).

\bibitem[{\citenamefont{Berthier et~al.}(2011)\citenamefont{Berthier, Biroli,
  Bouchaud, Cipelletti, and van Saarloos}}]{LucaDHbook}
\bibinfo{editor}{\bibfnamefont{L.}~\bibnamefont{Berthier}},
  \bibinfo{editor}{\bibfnamefont{G.}~\bibnamefont{Biroli}},
  \bibinfo{editor}{\bibfnamefont{J.-P.} \bibnamefont{Bouchaud}},
  \bibinfo{editor}{\bibfnamefont{L.}~\bibnamefont{Cipelletti}},
  \bibnamefont{and} \bibinfo{editor}{\bibfnamefont{W.}~\bibnamefont{van
  Saarloos}}, eds., \emph{\bibinfo{title}{Dynamical heterogeneities in glasses,
  colloids, and granular media}} (\bibinfo{publisher}{Oxford University Press},
  \bibinfo{year}{2011}).

\bibitem[{\citenamefont{Berthier}(2011)}]{BerthierP11}
\bibinfo{author}{\bibfnamefont{L.}~\bibnamefont{Berthier}},
  \bibinfo{journal}{Physics} \textbf{\bibinfo{volume}{4}}, \bibinfo{pages}{42}
  (\bibinfo{year}{2011}).

\bibitem[{\citenamefont{Katsuragi et~al.}(2010)\citenamefont{Katsuragi, Abate,
  and Durian}}]{katsuragi10}
\bibinfo{author}{\bibfnamefont{H.}~\bibnamefont{Katsuragi}},
  \bibinfo{author}{\bibfnamefont{A.~R.} \bibnamefont{Abate}}, \bibnamefont{and}
  \bibinfo{author}{\bibfnamefont{D.~J.} \bibnamefont{Durian}},
  \bibinfo{journal}{Soft Matter} \textbf{\bibinfo{volume}{6}},
  \bibinfo{pages}{3023} (\bibinfo{year}{2010}).

\bibitem[{\citenamefont{Nordstrom et~al.}(2011)\citenamefont{Nordstrom, Gollub,
  and Durian}}]{KerstinDH}
\bibinfo{author}{\bibfnamefont{K.~N.} \bibnamefont{Nordstrom}},
  \bibinfo{author}{\bibfnamefont{J.~P.} \bibnamefont{Gollub}},
  \bibnamefont{and} \bibinfo{author}{\bibfnamefont{D.~J.}
  \bibnamefont{Durian}}, \bibinfo{journal}{Phys. Rev. E}
  \textbf{\bibinfo{volume}{84}}, \bibinfo{pages}{021403}
  (\bibinfo{year}{2011}).

\bibitem[{\citenamefont{Elmatad et~al.}(2009)\citenamefont{Elmatad, Chandler,
  and Garrahan}}]{ECG}
\bibinfo{author}{\bibfnamefont{Y.~S.} \bibnamefont{Elmatad}},
  \bibinfo{author}{\bibfnamefont{D.}~\bibnamefont{Chandler}}, \bibnamefont{and}
  \bibinfo{author}{\bibfnamefont{J.~P.} \bibnamefont{Garrahan}},
  \bibinfo{journal}{J. Phys. Chem. B} \textbf{\bibinfo{volume}{113}},
  \bibinfo{pages}{5563} (\bibinfo{year}{2009}).

\bibitem[{\citenamefont{Haxton}(2011)}]{Haxton11}
\bibinfo{author}{\bibfnamefont{T.~K.} \bibnamefont{Haxton}},
  \bibinfo{journal}{arXiv:1103.3704}  (\bibinfo{year}{2011}).

\bibitem[{\citenamefont{O'Hern et~al.}(2003)\citenamefont{O'Hern, Silbert, Liu,
  and Nagel}}]{OHernPRE03}
\bibinfo{author}{\bibfnamefont{C.~S.} \bibnamefont{O'Hern}},
  \bibinfo{author}{\bibfnamefont{L.~E.} \bibnamefont{Silbert}},
  \bibinfo{author}{\bibfnamefont{A.~J.} \bibnamefont{Liu}}, \bibnamefont{and}
  \bibinfo{author}{\bibfnamefont{S.~R.} \bibnamefont{Nagel}},
  \bibinfo{journal}{Phys. Rev. E} \textbf{\bibinfo{volume}{68}},
  \bibinfo{pages}{011306} (\bibinfo{year}{2003}).

\bibitem[{\citenamefont{Liu and Nagel}(2010)}]{LiuNagelARCMP10}
\bibinfo{author}{\bibfnamefont{A.~J.} \bibnamefont{Liu}} \bibnamefont{and}
  \bibinfo{author}{\bibfnamefont{S.~R.} \bibnamefont{Nagel}},
  \bibinfo{journal}{Ann. Rev. Cond. Matt. Phys.} \textbf{\bibinfo{volume}{1}},
  \bibinfo{pages}{347} (\bibinfo{year}{2010}).

\end{thebibliography}

\end{document}